\documentclass[reprint,onecolumn,longbibliography,eqsecnum,floats,showpacs,%
nofootinbib,amsmath,amssymb,aps,prd,superscriptaddress,notitlepage]{revtex4-1}

\usepackage{amsmath,amssymb}
\usepackage{enumerate}

\usepackage[active]{srcltx}

\begin{document}

\newcommand{\re}{\mathbb{R}}
\newcommand{\integ}{\mathbb{Z}}
\newcommand{\compl}{\mathbb{C}}
\newcommand{\rd}{{\rm d}}
\newcommand{\lp}{\ell_{{\rm Pl}}}
\newcommand{\id}{\mathbb{I}}
\newcommand{\lat}{\mathcal{L}}
\newcommand{\Sp}{{\rm Sp}}

\newcommand{\ket}[1]{|{#1}\rangle}
\newcommand{\Dom}{\mathcal{D}}

\newcommand{\Hil}{\mathcal{H}}
\newcommand{\phy}{{\rm phy}}
\newcommand{\kin}{{\rm kin}}
\newcommand{\Hilg}{\Hil_{{\rm gr}}}
\newcommand{\Hilk}{\Hil_{{\rm kin}}}

\newcommand{\Bohr}{{\rm Bohr}}
\newcommand{\Fou}{\mathcal{F}}

\newcommand{\ubar}[1]{\underline{#1}}
\newcommand{\ub}[1]{\underline{#1}}

%
%

\title{Universe's memory and spontaneous coherence in loop quantum cosmology}

\author{Tomasz Paw{\l}owski}
\email{tpawlow@fuw.edu.pl}
\affiliation{Instytut Fizyki Teoretycznej, Wydzia{\l} Fizyki, Uniwersytet Warszawski, \\
	Pasteura 5, 02-093 Warszawa, Poland, EU. }


\begin{abstract}
The quantum bounce a priori connects several (semi)classical epochs of Universe evolution, 
however determining if and how well the semiclassicality is preserved in this transition 
is highly nontrivial. We review the present state of knowledge in that regards in the 
isotropic sector of loop quantum cosmology. This knowledge is next extended by studies
of an isotropic universe admiting positive cosmological constant (featuring an infinite chain of 
large Universe epochs). It is also shown, that such universe always admits 
a semiclassical epoch thanks to spontaneous spontaneous coherence, provided it is semiclassical 
in certain constant of motion playing the role of energy.
\end{abstract}

\keywords{loop quantum cosmology, quantum bounce, semiclassicality, coherence}

\pacs{98.80.Qc, 04.60.Pp}

\maketitle


\section{Introduction}	

Over one and a half decade since its birth the field of loop quantum cosmology (LQC) has 
experienced tremendous progress \cite{b-livrev,as-rev,bcmb-rev}. One of the flag features 
of the models developed within the area is the so called \emph{quantum bounce} -- a high 
energy (Planck order) epoch of an universe evolution providing a deterministic connection 
between two low energy epochs (when given universe adheres to the rules of classical 
General Relativity) \cite{aps-prl}. 

Since the process of the bounce is of pure quantum nature, bearing in fact a lot of similarities 
with the scattering process, it is a-priori not obvious whether the universe which is semiclassical 
before the bounce will evolve into a semiclassical one after it. Conversely, the same hold 
in regards of the question whether the expanding post-bounce semiclassical universe (like the 
observed one) had necessarily a semiclassical past. The numerical studies which originally 
established the bounce have shown strong indication that the answer to both these questions 
is in the affirmative, however by their very nature these studies could only cover a tiny 
(non-generic) portion of the space of physical states. As a consequence, the problem of 
semiclassicality preservation across the bounce is far from trivial even in the simplest 
isotropic sector of the theory and for a time was an arena of disagreement between researchers
\cite{cs-recall, b-comm, cs-comm}. 

Over time, several groups addressed this issue using both analytical and numerical 
methods. At present it is established, that the semiclassicality is indeed preserved at least
in the isotropic sector of the theory for the models featuring single bounce. There are also 
strong indications that this feature carries to the anisotropic homogeneous sector. Here we 
present a short review of the results which finally led to this conclusion. 

It it important to note though, that the vast majority of these studies features a particular
model (Friedman-Robertson-Walker (FRW) universe admitting massless scalar field). Furthermore, 
up to date there are no substantial studies (in the considered aspect) of models featuring 
infinite chains of low energy epochs connected by a sequence of bounces. Since the universe 
admitting positive cosmological constant (which is a feature of our Universe established with 
quite strong observational evidence) falls within this category, it is critical that the results 
are extended to it. The second part of this article is dedicated to this issue. There, 
the preservation result is extended to the case of (again isotropic) FRW universe with massless 
scalar field and positive cosmological constant. 

The presence of an infinite chain of low energy epochs with generically varying (between epochs) 
semiclassicality properties leads to an interesting question: how generic the semiclassical sector
is within the whole physical Hilbert space? Will an arbitrarily quantum universe eventually admit a semiclassical epoch? This question is addressed in the last part of the article, again in the context 
of FRW universe with positive cosmological constant and masseless scalar field. We indeed show, 
that, due to a process known in quantum mechanics as spontaneous coherence, for a universe to admit a
semiclassical epoch it is enough that the quantum state respresenting it is semiclassical with respect
to an observable (scalar field momentum) representing a constant of motion.

Since the vast majority of the presented material regards the specific model of isotropic universe 
with massless scalar field with the notion of physical (time) evolution tied to this field we 
start with briefly introducing the details of this model.

\section{Isotropic sector of LQC}

The particular model we will focus on is the flat (isotropic) FRW universe with massless 
scalar field and cosmological constant. We will follow the specific Hamiltonian formulation and 
quantization procedure as specified in \cite{acs-aspects,pa-posL}. 

\subsection{The Hamiltonian formulation}\label{sec:ham}

Our starting point is the standard Einstein-Hilbert action of gravity coupled to matter 
(in our case massless scalar field) with partial gauge fixing using the (physically 
distinguished) foliation by homogeneity surfaces. In the chosen gauge the spacetime 
metric takes the well known form
\begin{equation}
	g = -N^2(t) \rd t^2 + a^2(t) q^o , 
\end{equation}
where $N$ is the lapse function, $a$ is a scale factor and $q^o$ is the fixed, positive definite, 
flat metric (constant in the co-moving coordinates) known as \emph{fiducial metric}.
The natural $3+1$ splitting is next employed in the transition to Hamiltonian formalism, however 
due to homogeneity and noncompactness of the slices the integrals representing the symplectic 
structure and Hamiltonian diverge. To build meaningful theory ont thus introduces an infrared 
regulator ---a cell $\mathcal{V}$ taken to be cubical with sides along co-moving coordinates--- 
and restricts all integrals to it. The actual physical theory is then expected to emerge in the 
regulator removal limit defined by expanding $\mathcal{V}$ to fill entire slice.

The treatment follows that of LQG where one uses triads instead of $3$-metric 
directly. In the case considered here one can again fix the gauge (triad orientation)
such that the triad is determined by a single configuration variable $v$ 
which encodes both the volume $V$ of the cell $\mathcal{V}$ (with respect to physical 
spatial metric $q:=a^2(t)q^o$) 
and the orientation of the triad
\begin{equation}\label{eq:v-def} 
	({\rm sgn}\, v) \,\, v = \frac{V}{2\pi\gamma\sqrt{\Delta}\lp^2} \equiv
	\frac{a^3V_o}{2\pi\gamma\sqrt{\Delta}\lp^2}	
\end{equation}
where $\gamma$ is the Barbero-Immirzi parameter of LQG, $\Delta =
4\pi\sqrt{3}\gamma\, \lp^2$ is the so called \emph{LQC area gap} (see the next 
subsection) and $V_o$ is the volume of $\mathcal{V}$
with respect to $q^o$. The canonically
conjugate momentum (denoted by $b$) is on classical solutions given by
\begin{equation} 
	b = \gamma \sqrt{\Delta} H \, \equiv \, \gamma\sqrt{\Delta} \frac{1}{a}\frac{\rd a}{\rd t}  
\end{equation}
where $H$ is the Hubble parameter and $t$ is the proper (or cosmological) time. 

For the scalar field, the basic canonical pair is the standard one $\phi, p_{\phi}$. 
The total phase space is thus topologically $\mathbb{R}^4$ and equipped 
with basic Poisson brackets:
\begin{equation} 
	\{b,\, v\} = \frac{2}{\hbar}, \qquad \{\phi,p_{\phi}\} =1\, . 
\end{equation}

Following the standard procedure of building the Hamiltonian applied in LQG \cite{t-lqg} 
one arrives to reduced algebra of constraints, however due to gauge choice specified 
earlier the only nontrivial generator of this algebra is the Hamiltonian constraint:
\begin{equation}\label{eq:Cgr-wdw}
  {C} = \, p_{\phi}^2 - 3\pi\hbar^2 G b^2v^2 +
	\pi \gamma^2\Delta\, \hbar^2 G\, \Lambda\, v^2\,  \approx\, 0 ,
\end{equation}
Since there is no explicit dependence on $\phi$ there, the mnomentum $p_{\phi}$ 
is a constant of motion. This in turn implies that on the dynamical trajectory the field 
$\phi(t)$ is monotoneous function of cosmic time, thus is a good choice for an internal 
clock parametrising the evolution. In order to directly tie the Hamiltonian time to this field 
(which can be thought of as a kind of deprametrization procedure) we select the lapse 
function to be $N = a^3$ (thus following the so called SLQC prescription \cite{acs-aspects}).

\subsection{LQC quantization}
\label{sec:lqc}

The loop quantization procedure of the specified model is presented in detail 
in \cite{pa-posL} (which in turn follows the techniques of \cite{aps-imp,acs-aspects} and 
\cite{abl-lqc}) and is an adaptation of the procedure used for full LQG, implementing the 
Dirac program of quantizing the theories with constraints. This program involves the following
steps:
\begin{enumerate}[(i)]
	\item {\bf Kinematic level quantization:} 
		Building a suitable representation of the (reduced) holonomy-flux algebra through 
		GNS technique (whith GNS spectrum providing the Hilbert space).
	\item {\bf Building a quantum constraint operator:} In this case the Hamiltonian 
		constraint \eqref{eq:Cgr-wdw} is expressed in terms of holonomies and fluxed via procedure of Thiemann 
		regularization \cite{t-lqg}. In particular, following the heuristic argument of consistency of LQC 
		with full LQG the value of area gap  in \eqref{eq:v-def} is fixed as the lowest nonzero 
		eigenvalue of LQG area operator (although more intricate examination of LQC-LQG connection may lead to different choices \cite{p-interface}).
	\item {\bf Physical level quantization:} In this final step one constructs the physical Hilbert 
		space as kernel of quantum constraint operator and defines suitable algebra of Dirac 
		observables through partial observable formalism (essentially a family of constants of 
		motion parametrized by value of internal clock).
\end{enumerate}
As a result of $(i)$ one ends up with a kinematical Hilbert space
being a tensor product $\Hilk = \Hilg\otimes \Hil_{\phi}$, where 
$\Hil_{\phi} = L^2(\re, \rd \phi)$ and $\Hilg =
L^2(\bar{\re},\rd\mu_{\Bohr})$  (with $\bar{\re}$ being a Bohr
compactification of the real line and $\rd\mu_{\Bohr}$ the Haar
measure thereon). A convenient basis on $\Hilg$ is provided by the
eigenvectors of the operator $\hat{v}$:
\begin{equation}\label{eq:v-def-op}
  \hat{v}\ket{v} = v\ket{v}, \quad{\hbox{{\rm so that}}}\quad
  \hat{V}\ket{v}= (2\pi\gamma\sqrt{\Delta}\lp^2)\, |v|\, \ket{v}\, .
\end{equation}
In the volume representation states in $\Hilg$
become wave functions $\psi(v)$. To incorporate the fact that $v\to
-v$ is a large gauge transformation corresponding to the flip of the
orientation of the physical triad they are taken to be
symmetric $\psi(v) = \psi(-v)$. Unlike in standard quantum mechanics 
(Schr\"odinger representation) the $\psi(v)$ have support only on a 
countable set of points along the $v$-axis and their inner product is 
given by a sum
\begin{equation}\label{eq:ip-gr}
  \langle\psi|\psi'\rangle = \sum_{v\in\re}\,\, \bar{\psi}(v)
  \psi'(v)\, .
\end{equation}

The regularization of step $(ii)$ yields (after promoting the holonomies 
and fluxes invloved to operators) the quantum Hamiltonian constraint operator
of the form
\begin{equation}\label{eq:constr-quant}
  \hat{C} = \id\otimes\partial_{\phi}^2 + \Theta_{\Lambda} \otimes \id ,
  \qquad \Theta_{\Lambda} := \Theta_o - \pi G\gamma^2\lambda^2\,\Lambda\,
  v^2\, ,
\end{equation}
where
\begin{equation}\label{eq:ev2-gen}
-[\Theta_o\psi](v) = f_+(v)\,\psi(v-4) - f_o(v)\,\psi(v)+ f_-(v)\psi(v+4)\, ,
\end{equation}
with the coefficients $f_{o,\pm}$ given by
\begin{subequations}\label{eq:fslqc}\begin{align}
f_\pm(v) = (3\pi G/4)\, \sqrt{v(v\pm 4)}\,
	(v\pm 2), \qquad
		f_o(v) = (3\pi G/2) v^2 \, .
\end{align}\end{subequations}

The operator $\Theta_o$ is a second order difference
operator with uniform steps of size $v=\pm 4$ well defined on the domain $\Dom$ 
of finite linear combinations of $\ket{v}$.
Therefore, there is super-selection: one can investigate
dynamics separately on uniform lattices in the $v$-space and
each sector consisting of wave functions with support on any
one of these lattices is preserved by the complete set of Dirac
observables of interest. In this paper, we will restrict ourselves 
to the lattice $\lat = \{v=4n,\, n\in\integ\}$ for simplicity as 
in LQC physical results are largely insensitive to the choice of 
the sector \cite{mop-presc}.

For technical reasons it is more convenient to work in the dual 
representation in which states are wave functions $\psi(b)$ of 
the conjugate variable $b$. While operators corresponding to $b$ do not exist 
in the theory, mathematically one can define the transformation via 
Fourier series
\begin{equation}\label{eq:lqc-v-b-trans}
  [\Fou\psi](b) =
  \frac{1}{2\sqrt{\pi}}\, \sum_{\lat_0\setminus\{0\}} |v|^{-\frac{1}{2}}\,
  \psi(v)\, e^{\frac{i}{2}vb} ,
\end{equation}
where the point $v=0$ was removed from the transform because
the state with support just at $v=0$ is dynamically decoupled
from the orthogonal sub-space spanned by states which vanish at
$v=0$. Since $\psi$ are supported on $\lat_0$, their images
$\Fou\psi$ are periodic in $b$ with the period $\pi$. Therefore
one can restrict the support of the wave functions
$[\Fou\psi](b)$ just to the circle $b\in[0,\pi]$, with the
identification $[\Fou\psi](0)=[\Fou\psi](\pi)$.

By inspection, the elementary operators $\hat{v}$ and
$\hat{\mathcal{N}}_{\mu}$ defined by
\begin{equation} \label{ops}
  \hat{v}\ket{v} = v\ket{v} , \qquad {\rm and} \qquad
  \hat{\mathcal{N}}_{\mu}\ket{v} = \ket{v+\mu} ,
\end{equation}
in the $v$ representation are transformed to
\begin{equation}
  \hat{v} = 2i\partial_b , \qquad {\rm and} \qquad
  \mathcal{N}_{\mu} = e^{-i\mu b/2} .
\end{equation}
in the $b$ representation. As a consequence, the operator
$\Theta_{\Lambda}$ assumes the form
\begin{equation}\label{eq:TL-b}
  \Theta_{\Lambda} = - 12\pi G\,\, \big[\, (\sin(b)\partial_b)^2 -
  {\rm sgn}(\Lambda) b_o^2\partial_b^2 \big],
\end{equation}
in the $b$ representation, where $b_o:= \gamma\, \sqrt{\Lambda\Delta/3}$.

The properties of the operator $\Theta_{\Lambda}$ depend on the value (sign) 
of the cosmological constant, thus the last step in quantization program has 
to be performed for each case $\Lambda=0$, $\Lambda>0$, $\Lambda<0$ separately. 
Here we focus only on former two cases.

\subsubsection{$\Lambda=0$}\label{sec:L0}

The operator $\Theta_o$ is positive definite and essentially self-adjoint 
(thus generating a unique unitary evolution). Its spectrum $\Sp(\Theta_o) = \re^+$ is 
continuous and nondegenerate (on symmetric sector). Per analogy to Klein-Gordon equation
we restrict the physical Hilbert space to positive frequency states -- satisfying
\begin{equation}
  -i\partial_{\phi}\Psi(v,\phi) = \sqrt{\Theta_o}\Psi(v,\phi) . 
\end{equation}
Thus, the relevant Physical states are described by wave functions of the form
\begin{equation}\label{eq:L0-phy}
	\Psi(v,\phi) = \int_0^{+\infty} \rd k \tilde{\Psi}(k) e_k(v) e^{i\omega(k)\phi} , 
\end{equation}
where $\tilde{\Psi}$ is a \emph{spectral profile} of the wave function, the dispersion 
relation is $\omega(k) = \sqrt{12\pi G} k$ and $e_k$ are eigenbasis elements satisfying 
$\omega^2(k) e_k = \Theta_o e_k$ and normalized to satisfy 
$\langle e_k | e_{k'} \rangle = \delta(k-k')$. 

The physical inner product is
\begin{equation}
	\langle \Psi | \Phi \rangle = \int_0^{+\infty} \rd k \bar{\Psi}(k)\Phi(k) \ .
\end{equation}

As the physical observables it is convenient to select
\begin{enumerate}[(i)]
	\item The scalar field momentum (analog of energy in KG equation) operator
		\begin{equation}
			\hat{p}_{\phi} = \sqrt{\Theta_o},
		\end{equation}
	\item The family of 'volume at given $\phi$' operators defined by action
		\begin{equation}
			[ \hat{V}_{\phi_o} \Psi ](v,\phi) 
			= 2\pi\gamma\sqrt{\Delta}\lp^2 e^{i\Theta_o(\phi-\phi_o)} |v| \Psi(v,\phi_o) .
		\end{equation}
\end{enumerate}

\subsubsection{$\Lambda>0$}

The case of positive cosmological constant is a bit more complicated. Since the 
observations indicate the cosmological constant order of magnitude 
$\Lambda \sim 10^{-120} \lp^{-2}$ it is safe to assume $b_o<1$ in \eqref{eq:TL-b}. 
In this case $\Theta_{\Lambda}$ is no longer positive definite (although Hamiltonian 
constraint still selects out its positive part). Furthermore, it is no longer essentially 
self adjoint admitting instead a $1$-parameter family of selfadjoint extensions. Each of 
these extensions has purely discrete spectrum consisting of isolated points selected 
out by condition (with dispersion relation $\omega(k)= C_\omega k$, see \eqref{eq:Comega} 
for the value of $C_{\omega}$)
\begin{equation}\label{eq:SpL-cond}
	\tan(k_n y_o) + \tanh[k_n(\pi-y_o)] \tan(\beta) = 0 ,
\end{equation}
where $\beta\in [0,\pi)$ labels the extensions, the constant $y_o$ is expressed in terms
of the elliptic integral of the first kind
\begin{equation}\label{eq:yo-def}
	\pi y_o^{-1} := 1 + \frac{\sqrt{1-b_o^2}}{b_o} 
		\frac{F(\arcsin(b_o),1/b_o^2)}{F(\pi/2-\arcsin(b_o),1/(1-b_o^2))} ,
\end{equation}
and the proportionality constant in the dispersion relation is
\begin{subequations}\begin{align}
	C_\omega &:= \sqrt{12\pi G} \pi/x_M , \label{eq:Comega} \\
	x_M &:= \frac{1}{\sqrt{1-b_o^2}} F(\pi/2-\arcsin(b_o),1/(1-b_o^2)) 
	  + \frac{1}{b_o} F(\arcsin(b_o),1/b_o^2) .
\end{align}\end{subequations}
An important relation is the assymptotic behavior of $k_n$
\begin{equation}\label{eq:kn-ass}
  k_n = (n\pi -\beta)/y_o + O(e^{-2\pi n(\pi-y_o)/y_o}) .
\end{equation}
This relation will play a crucial role in obtaining the results of sections \ref{sec:DS-disp} 
and \ref{sec:DS-coh}.

The physical states have the form
\begin{equation}\label{eq:PsiL}
	\Psi(v,\phi) = \sum_{n=0}^{+\infty} \tilde{\Psi}_n e_{\beta,n}(v) e^{i\omega(k_n)\phi} , 
\end{equation}
where again $\tilde{\Psi}_n$ is (this time discrete) spectral wave function profile and 
$e_{\beta,n}$ are (explicitly) normalized eigenfunctions of the (positive part of the) 
extension of $\Theta_{\Lambda}$ corresponding to given value of $\beta$.

As the set of physical observables one can use analogs of the ones specified in section \ref{sec:L0},
however, as even classically the trajetory $V(\phi)$ reaches infinity for finite $\phi$ the 
operators $\hat{V}_{\phi}$ would not preserve the Hilbert space. Therefore one is forced to use their
compactified versions. Thus, finally we end up with
\begin{enumerate}[(i)]
	\item The ``energy'' operator
		\begin{equation}
			\hat{p}_{\phi} = \sqrt{|\Theta_{\Lambda}|},
		\end{equation}
	\item The family of 'compactified volume at given $\phi$' operators defined by action
		\begin{equation}\label{eq:Lp-obs}
			[ \hat{\theta}_{\phi_o} \Psi ](v,\phi) 
			= \theta_K(v) e^{i\Theta_o(\phi-\phi_o)} \Psi(v,\phi_o) ,
		\end{equation}
		where $\theta_K(v) := \arctan(|v|/K)$, with $K$ being positive constant of dimension of 
		the volume (of which particular value can be selected arbitrarily).
\end{enumerate}

\section{Preservation of semiclassicality: state of the art so far}
\label{sec:state}

As mentioned in the introduction the studies of semiclassicality focus almost entirely 
on the case of vanishing cosmological constant. There are essentially three lines 
of approach explored in the literature: $(i)$ analytical studies in manageable prescriptions, 
$(ii)$ numerical studies of selected classes of states, and $(iii)$ estimates following
from employing the interpretation of bounce as scattering. Let us start with the analytic approach.

\subsection{Solvable prescription of LQC: analytical results}

In general the loop quantization procedure presented in section \ref{sec:lqc} features 
a series of ambiguities. Various ways of fixing them lead to many prescriptions of LQC, 
several of which have been explored in the literature (see \cite{mop-presc} for their detailed 
comparative analysis). Most of them however require numerical methods to probe the state 
properties. The first prescription permitting reliable analytical treatment of the simplest case of 
universe with $\Lambda=0$ is known as \emph{Solvable LQC} \cite{acs-aspects} (this prescription 
is the one specified in section \ref{sec:lqc} and further used in studies of sections 
\ref{sec:DS-disp} and \ref{sec:DS-coh}). That prescription allowed to provide a strong estimate on 
the dispersion growth across the bounce for quite large class of states \cite{cs-recall}.

The key feature of the analysis was the fact, that (for $\Lambda=0$) upon switching to $b$ 
representation and further reparametrizing the ``momentum'' coordinate $b$ to a new one 
\begin{equation}
	x := 
	\ln(\tan(\sqrt{\Delta}b/2)) 
\end{equation}
one reduces the Hamiltonian constraint \eqref{eq:constr-quant} to explicit Klein-Gordon equation, 
which in turn yields the following form of physical states
\begin{equation}
	\Psi(x,\phi) = \int_0^{+\infty} \rd k \tilde{\Psi}(k) \cos(kx) e^{i\omega(k)\phi} ,
\end{equation}
with the dispersion relation $\omega(k)$ same as the one in sec.~\ref{sec:L0}. Moreover, 
upon defining a simple transformation from the physical Hilbert space to certain auxiliary one, 
the relevant observables also take a quite simple analytic form. This allows to parametrize 
the quantum trajectories (evolution of expectation values of the obsevables selected in 
sec.~\ref{sec:L0} and their dispersions) by a set of just $5$ parameters -- expectation values
of a set of operators corresponding to constants of motion \cite{acs-aspects}. This set of 
parameters captures in particular the information on how the dispersion grows across the bounce. 

The analysis of \cite{cs-recall} focuses of the states which at some moment (value of $\phi$) had
a support at $x\in[x_o-\epsilon,x_o+\epsilon]$ for certain large value $x_o$ (moment of evolution 
featuring low energy density) with $\epsilon\ll x_o$. For this class of states it was shown, that
\begin{equation}
	\left|[\lim_{\phi\to-\infty} - \lim_{\phi\to+\infty}](\Delta\hat{V}_{\phi}/\langle \hat{V} \rangle_{\phi})
	\right|
	\leq (1+\delta) (e ^{8\epsilon} -1) \sim 8\epsilon (1+\delta) ,
\end{equation}
where $\delta := \lim_{\phi\to+\infty}(\Delta\hat{V}_{\phi}/\langle \hat{V} \rangle_{\phi})$. 

Found inequality is exact for selected class of states, however the requirement 
of compactness of the support is quite restrictive and in general is believed to be too restrictive 
to admit large semiclassical sector (with respect to sufficiently large family of physically relevant observables).
The results of \cite{cs-recall} can be however extended beyond that family at the cost of becoming 
estimates rather than exact inequalities.

This method, although strong and precise, strongly relies on the ability to cast the studied model 
as a very simple one (Klein-Gordon system). This is possible only for a very narrow family of 
scenarios in isotropic LQC like flat FRW universe with dust \cite{hp-dust-LQC} (for any value 
of $\Lambda$) or with radiation \cite{ppwe-radiation} (for $\Lambda=0$) but so far has been 
impossible to extend even to the case of universe with massless scalar field and $\Lambda\neq 0$. 
Such scenarios require numerical analysis.

\subsection{Generalized Gaussian states: numerical studies}


In the pioneering work in which the bounce has been discovered as a feature of the model \cite{aps-imp} the
dynamics of quantum universe has been studied by purely numerical methods. The direct inspection 
of the large population of states have shown that its relative dispersions always satisfied the 
inequality
\begin{equation}\label{eq:ineq1}
	\left|\frac{\Delta\hat{V}_{\phi}}{\langle\hat{V}\rangle_{\phi}}
		-\frac{\Delta\hat{V}_{\pm}}{\langle\hat{V}\rangle_{\pm}}\right| 
		< \frac{\Delta\hat{p}_{\phi}}{\langle\hat{p}_{\phi}\rangle}
\end{equation}
throughout the evolution (that is for all probed values of $\phi$), where 
$\langle\hat{V}\rangle_{\pm} := \lim_{\phi\to\pm\infty}\langle\hat{V}\rangle_{\phi}$. 
In all the studied cases the states which started as semiclassical (sharply peaked 
in selected observables) at given initial $\phi=\phi_o$ remained so during the whole evolution.
The feature of semiclassicality preservation has been subsequently confirmed (by direct inspection) 
for all the cases of LQC dynamical evolution studied on the genuine quantum level: universe 
with spherical toopology \cite{apsv-spher}, with nonvanishing cosmological constant \cite{bp-negL, pa-posL} 
and with Maxwell 
field as matter content \cite{ppwe-radiation}.
However due to technical limitations in most of these cases the studies have been restricted 
to a finite number of examples corresponding to the Hamiltonian Gaussians, that is the states 
of spectral profile\footnote{In cases when the spectrum of the evolution generator was discrete the 
cutoff to the values of $k$ such that $\omega(k)$ was in its spectrum was taken.}
\begin{equation}
	\tilde{\Psi} = \frac{1}{\sqrt{\pi\sigma}} e^{-(k-k_o)^2/2\sigma^2} .
\end{equation}
Subsequently, in the case of flat FRW universe with massless scalar field and $\Lambda=0$ 
the preservation of the semiclassicality has been confirmed (heuristically, without explicit 
test of inequality \eqref{eq:ineq1} however with the observed increase of dispersion remaining 
within the same level of magnitude) for profiles different than Gaussian \cite{mop-presc,dgms-nong} but still 
for technical reasons the (necessarily numerical) studies have been restricted to few 
specific shapes only (i.e. Gaussians in $\ln(k)$, triangle ``sawtooth'' profiles).
Note, that for specifically tailored states the increase in the spread can be substantial \cite{dgs-chimera}, 
however this can happen only for the states which were never semiclassical during their whole evolution 
(usually very quantum in ``energy'' $p_{\phi}$).

The abovementioned results have been strengthened (again just in case of flat FRW universe with massless
scalar field and vanishing cosmological constant) in \cite{cm-semicl} where authors performed systematic 
numerical analysys of the parameter space of solutions corresponding to generalized Gaussian spectral 
profile, that is
\begin{equation}
	\tilde{\Psi}(k) = k^n e^{-\eta(k-\beta)^2} , \qquad \eta,\beta\in \compl, \ n\in\mathbb{N}
\end{equation}
Within this class of states it was shown, that once the physically resonable dispersion of a state is
selected in its asymptotic past (relative dispersions of the order of $10^{-60}$) the inequality \eqref{eq:ineq1} is strongly undersaturated.

The above results, while being strong, have been obtained just for specific nongeneric classes of states, 
thus can be taken only as the indication of the general property of semiclassicality preservation rather 
than its solid proof. To get a truly firm confirmation one needs to resort to methods which on the one hand 
are not restricted to just solvable prescriptions in LQC and on the other hand allow to reliably 
probe the properties of general physical states. One of such methods comes from treating the loop 
quantum cosmology evolution as a process of scattering of large semiclassical universe.

\subsection{The scattering picture of the bounce}

Loop quantization is not a unique route of quantizing the cosmological models. One can 
start with the reduced classical phase space and the Hamiltonian constraint \eqref{eq:Cgr-wdw}
and apply to it the standard methods of quantum mechanics (also following Dirac program parallel to LQC). 
This is known as Wheeler-DeWitt (WDW) 
quantization. The description it provides is usually much simpler than that of LQC, however 
(unless the modification due to exotic matter field is introduced) it fails to resolve the 
singularity. Indeed, the wave packets representing semiclassical universe hit the boundary 
representing classical singularity ($v=0$) which in turn generically introduces 
a nonuniqueness in the unitary evolution of a given universe. 

The principal case considered here: flat FRW universe with massless scalar field and 
vanishing cosmological constant has a very simple description in WDW approach: The kinematical
Hilbert space is the product $\ubar{\Hil}_{{\rm kin}} = L^2(\re,\rd v) \otimes L^2(\re,\rd\phi)$, 
the physical states are representing by wave function (decomposed in basis of generalized eigenvectors
of kinematical volume and scalar field operator)
\begin{equation}
	\ubar{\Psi}(v,\phi) = \int_{-\infty}^{+\infty} \tilde{\ubar{\Psi}}(k) \ubar{e}_k(v) e^{i\omega(k)\phi} ,
	\quad 
	\ubar{e}_k(v) = e^{ik\ln|v|} ,
\end{equation}
(with $\omega(k) = \sqrt{12\pi G}|k|$ and all the symbols in the equation being defined analogously to \eqref{eq:L0-phy})
and the observables can be defined analogously to LQC, as the analog of Schr\"odinger equation 
(in $\phi$ reparametrization) takes the form
$-i\partial_{\phi}\Psi(v,\phi) = 12\pi G \sqrt{|(v\partial_v)^2|} \Psi(v,\phi)$. 

An important feature of those LQC models where the universe can expand to infinite size is the existence
of their well defined WDW limit. In particular the model considered in sec.~\ref{sec:L0} features
the following large $v$ behavior of the basis functions
\begin{equation}
	e_k(v) = e^{i\alpha(k)} \ubar{e}_k(v) + e^{-i\alpha(k)} \ubar{e}_{-k}(v) + O(|v|^{-3}) .
\end{equation}
That allows to associate with each LQC state a contracting ($k>0$) and expanding ($k<0$) 
WDW ``limit'' state. For the states satisfying $\Delta\hat{p}_{\phi} <\infty$ the expectation values 
and dispersions of observable $\ln|\hat{v}|_{\phi}$ (defined analogously to \eqref{eq:v-def-op}) of LQC state
approach in the limit of $\phi\to\pm\infty$ the values of analogous observables of the corresponding
WDW (expanding/contracting respectively) limit states \cite{kp-scatter}. Thus, as long as we are 
interested in distant future/past only the global evolution of LQC state can be considered as a 
scattering process with a very simple scattering matrix 
\begin{equation}
	\langle k | \hat{\rho} | k' \rangle = e^{-2i\alpha(k)} \delta(k+k') .
\end{equation}
The hehavior of the phase rotation $\alpha(k)$ could be systematically analyzed via numerical methods, 
which in turn allowed to establish the following triange inequality true for every physical state with
finite dispersion in $\hat{p}_{\phi}$
\begin{equation}
	|\sigma_{+} - \sigma_{-}| < 2 \Delta\ln|\hat{p}_{\phi}/\hbar| , \qquad
	\sigma_{\pm} = \lim_{\phi\to\pm\infty} \Delta\ln|\hat{v}|_{\phi} .
\end{equation}
Preservation of semiclassicality is then a straightforward consequence of this inequality.

The scattering process privides not only general and exact result in considered case but also can be 
easily generalized to more complicated systems not treateble analytically. Its downside is that it only 
allows to probe the asymptotic values, in particular not teling anything about the properties of a universe in high curvature region near the bounce.


\section{Dispersion and semiclassicality of DeSitter universe}
\label{sec:DS-disp}

The scattering picture presented in last sub-section can be generalized with 
reasonable effort to the case of positive cosmological constant. The LQC quantization of this 
model as well as its WDW limit have been studied extensively in \cite{pa-posL}. In comparizon to 
$\Lambda=0$ the model exhibits two important differences:
\begin{enumerate}[(i)]
	\item The volume of the universe reaches infinity for finite value of scalar field. That 
		leads to nontrivial extensions past this point and to transition between expanding and contracting 
		epoch of evolution through deSitter timelike future/past SCRI. This is a feature of both LQC and WDW approach, thus, 
		unlike in $\Lambda=0$ case, the eigenfuctions of the WDW evolution generator also take the form 
		of standing waves.
	\item Due to above, the bounce leads to infinite chain of large size low curvature epochs (universes)
		connected by quantum bounces and SCRI transitions.
\end{enumerate}
Furthermore, the unitary evolution of the system is nonunique (in both LQC and WDW approach). The choice
of self-adjoint extension of the ``Hamiltonian'' $\sqrt{|\Theta_{\Lambda}|}$ corresponds to the choice 
of boundary conditions at the SCRI. From the point of view of semiclassicality preservation this 
nonuniqueness is not critical, as one can always work with a single extension (with discrete spectrum of 
the evolution generator).

Similarly to $\Lambda=0$ one could start with establishing the asymptotics between the eigenfunctions of 
respectively LQC and WDW evolution generators. However, since WDW eigenfunctions themselves are standing 
waves they too would have to be decomposed onto simpler ``expanding'' (incoming to SCRI) 
and ``contracting'' (outgoing from SCRI) components. 

In LQC approach the direct inspection (for detailed description of method used 
in the identification of the limit see \cite{kp-posL}) shows that
\begin{equation}\label{eq:LQC-lim}
	e_n^{\beta}(v) = N_n [ e^{i\alpha} e^+_n (v) + e^{-i\alpha} e^-_n (v) ] + O(v^{-3}) , \quad 
\end{equation}
where 
\begin{subequations}\begin{align}
  e^{\pm}_n &= |v|^{-1} \ e^{\pm i\Omega|v|} \cdot e^{\pm i\kappa(n,\Lambda,\beta)/|v|} \\
  \cos(4\Omega(\Lambda)) &= 1-2\Lambda/\Lambda_c =: 1-2\lambda, \label{eq:Omega-def} \\
  \kappa(n,\Lambda,\beta) 
		&= \frac{3\pi G(1-2\lambda) + \omega_n^2}{12\pi G\sqrt{\lambda(1-\lambda)}} 
		=: A\omega_n^2 + B , \label{eq:AB-def}
\end{align}\end{subequations}
and $N_n$ is a normalization constant.
Note that the function $\theta$ used in \eqref{eq:Lp-obs} behaves at large $v$ as follows
\begin{equation}
  1/|v| \approx (1/K)(\theta_a-\pi/2) . 
\end{equation}

Similarly, the WDW basis eigenfunctions (corresponding to possibly different value of 
cosmological constant here denoted as $\ubar{\Lambda}$) exhibit the limit
\begin{equation}\label{eq:WDW-lim}
	\ub{e}_k^{\beta}(v) = N(k) [ e^{i\alpha} \ub{e}^+_k (v) + e^{-i\alpha} \ub{e}^-_k (v) ] + O(v^{-3}) ,
\end{equation}
where the phase rotation $\alpha$ depends on both $\beta$ and $k$ and 
\begin{subequations}\begin{align} 
	\ub{e}^{\pm}_k &= |v|^{-1} \ e^{\pm i\Omega|v|} \cdot e^{\pm i\kappa(k,\Lambda,\beta)/|v|} \\
	\kappa &= \kappa(\omega=\sqrt{12\pi G}k) , 
\end{align}\end{subequations}
and $\Omega$ the same as in \eqref{eq:Omega-def}. We then observe that the functions $e_{k}^{\pm}$ and 
$\ubar{e}_{k}^{\pm}$ do agree, provided we choose
\begin{equation}
	\ub{\Lambda}/\Lambda_c = \ub{\lambda} = \arccos(1-2\lambda) .
\end{equation}
This allows us to associate with each LQC basis eigenfunction a WDW one. 
This in principle allows to define a WDW limit of LQC state (barring one caveat which we will 
discuss in detail below), however with WDW basis functions themselves being quite complicated exploiting this fact is not practical for the purpose of semiclassicality preservation analysis. Instead we will 
construct the auxiliary 
\emph{limit Hilbert spaces} directly from $e_{k}^{\pm}$. Let us denote them by $\Hil^{\pm}$ and $\ubar{\Hil}^{\pm}$ for LQC and WDW respectively. We equip them with inner products selected in such 
a way that the norm of each (component) limit determined through 
\eqref{eq:LQC-lim}, \eqref{eq:WDW-lim} agrees with the norm of the original (LQC/WDW) state. 
We note, that on each limit space the term $e^{\pm i\Omega |v|}$ is a global rotation, thus can 
be dropped. After this modification the bases of the limit spaces become regular in $\theta$. 


Let us return for a moment to relating the LQC and WDW states. We do this through the sequence of transformations 
\begin{equation}
	\Hil_{\phy} \to \Hil^{\pm} \to \ubar{\Hil}^{\pm} \to \ubar{\Hil}_{\phy} . 
\end{equation}
We note immediately however, that the direct association through basis finctions (as specified earlier) 
would lead to associating with any formalizable element of $\Hil_{\phy}$ the WDW state(s) of zero norm. 
Thus, to be meaningful, the transformation $\Hil^{\pm} \to \ubar{\Hil}^{\pm}$
has to be modified. In order to select the appropriate modification of it we recall, that we defined the 
auxiliary spaces and searched for the limit in order to determine the semiclassicality properties of 
LQC state. Thus, a natural requirement for the sesired transformation is that the physical parameters (expectation values and dispersions of relevant observables) of the localized state near the SCRI are well reflected by those of the limit state. We can then define an \emph{instantiation} of the state: 
require an agreement of the relevant observables at time $\phi=\phi_o$. 
A priori one could use for that the so-called Hamburger \cite{bs-eff} decomposition and 
require that all the Hamburger moments are preserved upon the transformation. This approach however, 
although precise is impractical as reproducing the state (wave function) out of its Hamburger moments is extremely difficult and till now remains an open problem. 

Instead, we propose a simpler (although non-unique) construction motivated by properties of $1$-dimensional 
Klein-Gordon equation. We note that the same problem occurrs if for the system of free particle between two parallel walls we want to associate with a given particle (say at the moment of reflection from the wall) a wave packet
of a particle moving freely on $\re$. To define an instantiation at $\phi=\phi_o$ we thus follow the construction natural for that scenario.
\begin{enumerate}
	\item First by transformation of the spectral profile 
		$\tilde{\Psi}(k)\mapsto\tilde{\Psi}(k) e^{i\omega(k)\phi_o}$ 
		we reduce the problem to constructing the instantiation at $\phi=0$. 
	\item We extend the spectral profile from the discrete set of $k_n$ by linear interpolation 
		\emph{of the modulus and phase of $\tilde{\Psi}$ separately}.
	\item We transform the new wave function back using the inverse of the first step.
\end{enumerate}
Have the spectrum of $\Theta_{\Lambda}$ been uniform, this procedure would lead to definition 
of the WDW state of the same expectation value and dispersion of $\hat{p}_{\phi}$ as those of 
the original LQC state. However, it is almost regular for large $k$. Furthermore
to be semiclassical the states have to be peaked at large $k$ where the estimate \eqref{eq:kn-ass} 
is extremely accurate (the deviation from uniformity can be bounded by $C\cdot \exp(-2\pi n(\pi-y_o)/y_o)$
where $C$ is of the order of $1$). As a consequence, for a state of 
$\delta\hat{p}_{\phi}/\langle\hat{p}_{\phi}\rangle < \epsilon$ (where $\epsilon\ll 1$) the deviation
between expectation values and dispersions of $\ln|\hat{p}_{\phi}/\hbar|$ will be of the same order
: $C\cdot \exp(-2(1-\epsilon) \pi n(\pi-y_o)/y_o)$.

In order to determine how the state disperses through the bounce we need to compare the dispersions
of observable $\hat{\theta}$ at two consecutive moments of reflection from SCRI. Due to complicated 
form of the basis functions $e_n^\beta (v)$ doing so directly on $\Hil_{\phy}$ is extremely difficult.
Therefore, we cast the problem as comparing the relevant dispersions of the analog of $\hat{\theta}$ 
between two consecutive instantiations (corresponding to the points where the expectation value of 
$\hat{\theta}$ reaches maximum) on the auxiliary Hilbert space $\ub{\Hil}^+$. 
This analog observable takes at $\phi=0$ a very simple form 
\begin{equation}\label{eq:obs-x}
	\hat{\theta}^+ =: \hat{x} +\frac{\pi}{2}\id = \frac{iK}{2A\omega}\partial_{\omega} 
		+\frac{\pi}{2}\id ,
\end{equation}
where $A$ has been defined in \eqref{eq:AB-def}.

The main limitation of this step is a direct consequence of the fact, that even at the point where universe 
reaches SCRI $\Delta\hat{\theta}$ remains finite, thus even for sharply peaked states there would 
be finite differences between the dispersion of the original observable and its analog on the auxiliary 
space -- there is no exact convergence as observed in the case of $\Lambda=0$. The form of asymptotics 
\eqref{eq:LQC-lim} and the numerical observations of the behavior of $e_n^\beta$ allow to conclude, 
that for as long as $\Delta\hat{\theta}$ remains small the difference is of higher order, thus one can 
provide estimate 
\begin{equation}\label{eq:conj}
	1/C_\theta < \Delta\hat{\theta} / \Delta\hat{x} < C_\theta ,
\end{equation}
where $C_\theta$ is of the order of $1$. The sufficiently optimal value of $C_\theta$ and the domain 
of validity of the above estimate (corresponding to such value) can be determined precisely via numerical analysis. For the purpose of studies of this article we take it as a conjecture.

To estimate the difference of $\hat{x}$ between specified instantiations we note, that, have the spectrum of $\Theta_{\Lambda}$ been uniform, the 
values of interest would have agreed exactly. By the same argument as used in case of 
$\hat{p}_{\phi}$ we conclude that the relative change of dispersion of $\hat{x}_{\phi}$ is of the order 
$C\cdot \exp(-2(1-\epsilon) \pi n(\pi-y_o)/y_o)$. Thus for physically relevant semiclassical states 
they remain extremely small. As a consequence the semiclassicality is preserved at least in the sense 
of limiting states and by conjecture \eqref{eq:conj} we can extend it to physical states.

An important property of the studied system follows from the fact that the spectrum of $\Theta_{\Lambda}$ does deviate from uniformity, resulting in a nontrivial spread of the semiclassical wave packet which \emph{does occurr} across the bounce. In consequence, after sufficiently large (although enormous) number of cycles the originally semiclassical state will eventually loose the semiclassicality. As a consequence even the universe semiclassical at some moment of evolution will preserve this property only for finite time (although spanning many evolution cycles) which in turn may render semiclassicality a non-generic feature even for a single (dynamical trajectory of a) universe. To check whether this is indeed the case we can ask a converse question: given a generic quantum state, under what condition it will ever admit a semiclassical epoch? To answering that question we dedicate the next section.

\section{Spontaneous coherence in LQC}
\label{sec:DS-coh}

The process of the (originally dispersed) quantum state attanining in the process of dynamical 
evolution semiclassical properties is a feature of several quantum mechanical systems and known as \emph{spontaneous coherence}. The simple and regular structure of isotropic quantum cosmological 
models allows to expect that such process will occurr also in isotropic sector of LQC. In this section 
we investigate this process in context of (again) flat FRW universe with massless scalar field and positive cosmological constant.

To start with, we note that, since $p_{\phi}$ is a constant of motion, the semiclassicality properties 
tied to observable $\hat{p}_{\phi}$ cannot change, thus small relative dispersion with respect to this observable is a necessary condition for the state to be able to ever feature semiclassical epoch.

To probe the spread in volume we employ the tools developed in previous section, casting the problem 
as the issue of coherence of observable $\hat{x}_{\phi}$ \eqref{eq:obs-x} on the instantiation of LQC 
state.

Since the limiting states bear some similarity to plane Klein-Gordon waves the question of coherence can be posed as 
question about the existence of epoch when the Heisenberg unceratainty principle is close to be saturated
for the instantiated limiting wave packet.

Due to the chosen construction of instantiation (linear interpolation in phases) one can estimate the dispersion in $\hat{x}$ by differences in phases of (instantiated) $\tilde{\Psi}(k)$. The Heisenberg 
uncertainty takes the form
\begin{equation}\label{eq:Hei}
	\delta\hat{p}_{\phi} \Delta\hat{x}_{\phi} \geq \sqrt{12\pi G} \langle \hat{p}_{\phi}\rangle/2
\end{equation}

Consider now the state peaked at frequency $\omega_o$ and such that 
$\Delta\hat{p}_{\phi}/\langle\hat{p}_{\phi}\rangle<\epsilon_p\ll 1$.
If the phase of instantiated profile $\tilde{\Psi}(k)$ is severely bounded: $\tilde{\Psi}(k) \in [-\epsilon_{\varphi},\epsilon_{\varphi}]$ then the form of $\hat{x}$ \eqref{eq:obs-x} allows for the estimate
\begin{equation}\label{eq:eps}
	\Delta\hat{x}_{\phi=0} \leq C_x \cdot \frac{2 \epsilon_{\varphi}}{\omega_o}
\end{equation}
where $C_x$ is of the order of one.

We note, that all the information about the instantiated state is still contained in the cutoff of the 
profile $\tilde{\Psi}(k)$ to the original $k_n$ corresponding to sepctrum of $\Theta_{\Lambda}$. 
We then can encode the dynamical evolution as rotation of the spectral profile by phases 
$e^{i\omega(k_n)\phi}$ (with again linear interpolation between the discrete points as defined 
for the instantiations). The observable $x$ can then be always evaluated at $\phi=0$. 

Since the state is localized in $p_{\phi}$ (that is 
$\Delta\hat{p}_{\phi}/ \epsilon_p\hat{p}_{\phi}\rangle\leq \epsilon_p \ll 1 $) we can restrict the spectral profile to the finite number of points within the interval $\omega_n \in \omega_o[1-3\epsilon_\omega,1+3\epsilon_\omega]$, where $\epsilon_p\ll\epsilon_\omega\ll 1$. 
Indeed, the neglected part of the wave function has a norm necesarily smaller than $\epsilon_p/(3\epsilon_\omega)$, which allows in turn to estimate the correction to $\Delta\hat{x}$ 
via \eqref{eq:obs-x} as $K/(2A\omega_o\Delta\omega) \cdot \epsilon_p/(3\epsilon_\omega)$. 
Upon this restriction the evolution of the phases of the state is a smooth trajectory 
on $N$-dimensional torus (where $N$ is the number of eigenvalues in the selected interval). 

Since now the system is reduced to a finite one, we can use number theory to estimate how big $\epsilon_{\varphi}$ has to be in order for the trajectory to eventually hit a cell centered 
at $\vec{0}$ and of size $2\epsilon_{\varphi}$. This is a exxtbook ergodicity problem for a
linear dynamical systems on $n$-torus.
We observe, that the density of trajectory on the
torus depends on whether pairs of frquencies are rationally related. If none of frequencies $\omega_n$ are 
rationally related then the trajectory is (truly) dense on $N$-torus, thus $\epsilon_{\varphi}$ can be arbitrarily small. Therefore we can divide the selected set of frequencies (eigenvalues) onto classes 
of rationally related ones and consider each class separately. Within each class, in order to ensure 
that the trajectory intersects the distinguished cell one needs to select the size of the cell such that
\begin{equation}
	2\pi /\epsilon_\phi \leq \min_{(m,n)\in P_k} q_{m,n} ,  
\end{equation}
where $P_k$ is a set of all possible pairs of eigenvalues such that $\omega_m<\omega_n$ 
and the ratios of eigenvalues within each pair have the reduced form 
$\omega_m/\omega_n = p_{m,n}/q_{m,n}$ with $p_{m,n},q_{m,n}\in\mathbb{N}$.

While we cannot determine the relevant minimum, the fact, that the spectrum approaches uniformity extremely fast (while not being exactly uniform) provides us with the lower bound for it. Indeed \eqref{eq:kn-ass} immediately implies that the (reduced) denominator is higher than
\begin{equation}
	e^{2(1-3\epsilon_p) \pi n (\pi-y_o)/y_o} .
\end{equation}
As a consequence one can safely take
\begin{equation}
	\epsilon_{\phi} \sim 4\pi e^{-2(1-3\epsilon_p) \pi n (\pi-y_o)/y_o} .
\end{equation}
Together with \eqref{eq:eps} this estimate ensures existence of an epoch where the product $\delta\hat{p}_{\phi} \Delta\hat{x}_{\phi}$ is of the order of its minimal value allowed by uncertainty principle (see \eqref{eq:Hei}). Thus by the conjecture \eqref{eq:conj} we conclude that \emph{every} 
state sharply peaked in $\hat{p}_{\phi}$ admits semiclassical epoch in its dynamical evolution. Furthermore, 
finite size of $\epsilon_{\phi}$ ensures that the semiclasical epoch will be always hit after final 
time, thus excluding the possibility that semiclassical epoch is nongeneric along a single 
universe dynamical trajectory.

\section{Perspectives}

To summarize the original research reported in this article, we extended the known results regarding 
semiclassicality preservation of isotropic universe within the LQC framework to the case of DeSitter 
FRW universe admitting massless scalar field as the sole matter content. In such case:
\begin{enumerate}
	\item The semiclassical universe remains sharply peaked through many cycles of the evolution 
		(separated by quantum bounce), although it very slowly looses its semiclassicality properties, and
	\item it is enough that the universe is semiclassical with respect to the constant of motion $p_{\phi}$
		to admit a semiclassical epoch somewhere along its dynamical evolution.
\end{enumerate}
These results, together with existing ones reviewed in sec.~\ref{sec:state}, while promissing, 
are restricted to just several models within the isotropic sector of LQC. 
In order to be considered reliable they have to be extended not only to wider class of isotropic systems
(different topology and matter content) but also beyond the class of isotropic models: to homogeneous 
non-isootropic (like for example Bianchi I \cite{awe-b1}) and ultimately inhomogeneous scenarios.

Within the homogeneous sector of LQC most of the methods presented in this article: scattering picture, 
numerical analysis of generalized Gaussians as well as methods applied in sec~\ref{sec:DS-disp} and~\ref{sec:DS-coh} can be applied to wide class of the models\footnote{While in sec~\ref{sec:DS-disp} and~\ref{sec:DS-coh} the studies used the analytically determined spectrum of the evolution operator, the
methods applied there do not loose their efficiency if the spectrum has to be determined numerically as only estimates/bounds are needed.}. 
Beyond the homogeneity the situation complicated significantly. There, two particular approaches give 
a hope of success:
\begin{enumerate}[1.]
	\item The so called \emph{Abelianization procedure} \cite{gop-abel} when applied to cosmological models
		(see in particular an application to Gowdy cosmology \cite{mop-abel}) allows to bring the (otherwise 
		unbearably complicated) evolution generator to the form quite similar to the one known from
		isotropic sector of LQC. Provided, that the method is improved to unambiguously reproduce general relativity as its low energy limit, that property in principle allows to apply the methods discussed in this article either directly or after an extension.
	\item While the present studies via either dressed \cite{aan-pert3} or rainbow \cite{adl-rainbow} metric start to include 
		the effects of quantum dispersion of the states, up to now they do not allow for precise control of 
		the dispersion's behavior. One may however hope that the synthesis of above methods with the semiclassical
		approach using the Hamburger decomposition of the state \cite{bs-eff} will give birth to a methodology of 
		probing the universe dynamics sufficiently robust to address the semiclassicality loss or spontaneous coherence problems in realistic cosmological scenarios.
\end{enumerate}

Author thanks Parampreet Singh for helpful comments. This work has been supported in part by the Polish Narodowe Centrum Nauki (NCN) grant 2012/05/E/ST2/03308 and the Chilean grant\\ 
CONICYT/FONDECYT/REGULAR/1140335. 

\bibliographystyle{JHEP}
\bibliography{p-ijmpd}

\providecommand{\href}[2]{#2}\begingroup\raggedright\begin{thebibliography}{10}

\bibitem{b-livrev}
M.~Bojowald, {\it {Loop quantum cosmology}},  {\em Liv.Rev.Rel.} {\bf 11}
  (2008) 4.

\bibitem{as-rev}
A.~Ashtekar and P.~Singh, {\it {Loop Quantum Cosmology: A Status Report}},
  {\em Class.Quant.Grav.} {\bf 28} (2011) 213001,
  [\href{http://xxx.lanl.gov/abs/1108.0893}{{\tt arXiv:1108.0893}}].

\bibitem{bcmb-rev}
K.~Banerjee, G.~Calcagni, and M.~Mart\'in-Benito, {\it {Introduction to loop
  quantum cosmology}},  {\em SIGMA} {\bf 8} (2012) 016,
  [\href{http://xxx.lanl.gov/abs/1109.6801}{{\tt arXiv:1109.6801}}].

\bibitem{aps-prl}
A.~Ashtekar, T.~Paw{\l}owski, and P.~Singh, {\it {Quantum nature of the big
  bang}},  {\em Phys.Rev.Lett.} {\bf 96} (2006) 141301,
  [\href{http://xxx.lanl.gov/abs/gr-qc/0602086}{{\tt gr-qc/0602086}}].

\bibitem{cs-recall}
A.~Corichi and P.~Singh, {\it {Quantum bounce and cosmic recall}},  {\em Phys.
  Rev. Lett.} {\bf 100} (2008) 161302,
  [\href{http://xxx.lanl.gov/abs/0710.4543}{{\tt arXiv:0710.4543}}].

\bibitem{b-comm}
M.~Bojowald, {\it {Comment on `Quantum bounce and cosmic recall' PUB-NOTE =
  Phys.Rev.Lett.101:209001,2008}},  {\em Phys. Rev. Lett.} {\bf 101} (2008)
  209001, [\href{http://xxx.lanl.gov/abs/0811.2790}{{\tt arXiv:0811.2790}}].

\bibitem{cs-comm}
A.~Corichi and P.~Singh, {\it {Reply to `Comment on `Quantum Bounce and Cosmic
  Recall''}},  {\em Phys. Rev. Lett.} {\bf 101} (2008) 209002,
  [\href{http://xxx.lanl.gov/abs/0811.2983}{{\tt arXiv:0811.2983}}].

\bibitem{acs-aspects}
A.~Ashtekar, A.~Corichi, and P.~Singh, {\it {On the robustness of key features
  of loop quantum cosmology}},  {\em Phys.Rev.} {\bf D77} (2008) 024046,
  [\href{http://xxx.lanl.gov/abs/0710.3565}{{\tt arXiv:0710.3565}}].

\bibitem{pa-posL}
T.~Paw{\l}owski and A.~Ashtekar, {\it {Positive cosmological constant in loop
  quantum cosmology}},  {\em Phys.Rev.} {\bf D85} (2012) 064001,
  [\href{http://xxx.lanl.gov/abs/1112.0360}{{\tt arXiv:1112.0360}}].

\bibitem{t-lqg}
T.~Thiemann, {\em {Modern canonical quantum general relativity}}.
\newblock Cambridge University Press, London, 2007.

\bibitem{aps-imp}
A.~Ashtekar, T.~Paw{\l}owski, and P.~Singh, {\it {Quantum nature of the big
  bang: Improved dynamics}},  {\em Phys.Rev.} {\bf D74} (2006) 084003,
  [\href{http://xxx.lanl.gov/abs/gr-qc/0607039}{{\tt gr-qc/0607039}}].

\bibitem{abl-lqc}
A.~Ashtekar, M.~Bojowald, and J.~Lewandowski, {\it {Mathematical structure of
  loop quantum cosmology}},  {\em Adv.Theor.Math.Phys.} {\bf 7} (2003)
  233--268, [\href{http://xxx.lanl.gov/abs/gr-qc/0304074}{{\tt
  gr-qc/0304074}}].

\bibitem{p-interface}
T.~Paw{\l}owski, {\it {Observations on interfacing loop quantum gravity with
  cosmology}},  {\em Phys. Rev.} {\bf D92} (2015), no.~12 124020,
  [\href{http://xxx.lanl.gov/abs/1411.0323}{{\tt arXiv:1411.0323}}].

\bibitem{mop-presc}
G.~A. Mena~Marug\'an, J.~Olmedo, and T.~Paw{\l}owski, {\it {Prescriptions in
  Loop Quantum Cosmology: A comparative analysis}},  {\em Phys. Rev.} {\bf D84}
  (2011) 064012, [\href{http://xxx.lanl.gov/abs/1108.0829}{{\tt
  arXiv:1108.0829}}].

\bibitem{hp-dust-LQC}
V.~Husain and T.~Paw{\l}owski, {\it {Dust reference frame in quantum
  cosmology}},  {\em Class. Quant. Grav.} {\bf 28} (2011) 225014,
  [\href{http://xxx.lanl.gov/abs/1108.1147}{{\tt arXiv:1108.1147}}].

\bibitem{ppwe-radiation}
T.~Paw{\l}owski, R.~Pierini, and E.~Wilson-Ewing, {\it {Loop quantum cosmology
  of a radiation-dominated flat FLRW universe}},
  \href{http://xxx.lanl.gov/abs/1404.4036}{{\tt arXiv:1404.4036}}.

\bibitem{apsv-spher}
A.~Ashtekar, T.~Paw{\l}owski, P.~Singh, and K.~Vandersloot, {\it {Loop quantum
  cosmology of k=1 FRW models}},  {\em Phys.Rev.} {\bf D75} (2007) 024035,
  [\href{http://xxx.lanl.gov/abs/gr-qc/0612104}{{\tt gr-qc/0612104}}].

\bibitem{bp-negL}
E.~Bentivegna and T.~Paw{\l}owski, {\it {Anti-deSitter universe dynamics in
  LQC}},  {\em Phys.Rev.} {\bf D77} (2008) 124025,
  [\href{http://xxx.lanl.gov/abs/0803.4446}{{\tt arXiv:0803.4446}}].

\bibitem{dgms-nong}
P.~Diener, B.~Gupt, M.~Megevand, and P.~Singh, {\it {Numerical evolution of
  squeezed and non-Gaussian states in loop quantum cosmology}},  {\em Class.
  Quant. Grav.} {\bf 31} (2014) 165006,
  [\href{http://xxx.lanl.gov/abs/1406.1486}{{\tt arXiv:1406.1486}}].

\bibitem{dgs-chimera}
P.~Diener, B.~Gupt, and P.~Singh, {\it {Chimera: A hybrid approach to numerical
  loop quantum cosmology}},  {\em Class. Quant. Grav.} {\bf 31} (2014) 025013,
  [\href{http://xxx.lanl.gov/abs/1310.4795}{{\tt arXiv:1310.4795}}].

\bibitem{cm-semicl}
A.~Corichi and E.~Montoya, {\it {Coherent semiclassical states for loop quantum
  cosmology}},  {\em Phys. Rev.} {\bf D84} (2011) 044021,
  [\href{http://xxx.lanl.gov/abs/1105.5081}{{\tt arXiv:1105.5081}}].

\bibitem{kp-scatter}
W.~Kami{\'n}ski and T.~Paw{\l}owski, {\it {Cosmic recall and the scattering
  picture of Loop Quantum Cosmology}},  {\em Phys.Rev.} {\bf D81} (2010)
  084027, [\href{http://xxx.lanl.gov/abs/1001.2663}{{\tt arXiv:1001.2663}}].

\bibitem{kp-posL}
W.~Kaminski and T.~Pawlowski, {\it {The LQC evolution operator of FRW universe
  with positive cosmological constant}},  {\em Phys. Rev.} {\bf D81} (2010)
  024014, [\href{http://xxx.lanl.gov/abs/0912.0162}{{\tt arXiv:0912.0162}}].

\bibitem{bs-eff}
M.~Bojowald, H.~Hernandez, and A.~Skirzewski, {\it {Effective equations for
  isotropic quantum cosmology including matter}},  {\em Phys. Rev.} {\bf D76}
  (2007) 063511, [\href{http://xxx.lanl.gov/abs/0706.1057}{{\tt
  arXiv:0706.1057}}].

\bibitem{awe-b1}
A.~Ashtekar and E.~Wilson-Ewing, {\it {Loop quantum cosmology of Bianchi I
  models}},  {\em Phys.Rev.} {\bf D79} (2009) 083535,
  [\href{http://xxx.lanl.gov/abs/0903.3397}{{\tt arXiv:0903.3397}}].

\bibitem{gop-abel}
R.~Gambini, J.~Olmedo, and J.~Pullin, {\it {Quantum black holes in Loop Quantum
  Gravity}},  {\em Class. Quant. Grav.} {\bf 31} (2014) 095009,
  [\href{http://xxx.lanl.gov/abs/1310.5996}{{\tt arXiv:1310.5996}}].

\bibitem{mop-abel}
D.~M. de~Blas, J.~Olmedo, and T.~Paw{\l}owski, {\it {Loop quantization of the
  Gowdy model with local rotational symmetry}},
  \href{http://xxx.lanl.gov/abs/1509.0919}{{\tt arXiv:1509.0919}}.

\bibitem{aan-pert3}
I.~Agullo, A.~Ashtekar, and W.~Nelson, {\it {The pre-inflationary dynamics of
  loop quantum cosmology: Confronting quantum gravity with observations}},
  {\em Class.Quant.Grav.} {\bf 30} (2013) 085014,
  [\href{http://xxx.lanl.gov/abs/1302.0254}{{\tt arXiv:1302.0254}}].

\bibitem{adl-rainbow}
M.~Assanioussi, A.~Dapor, and J.~Lewandowski, {\it {Rainbow metric from quantum
  gravity}},  {\em Phys. Lett.} {\bf B751} (2015) 302--305,
  [\href{http://xxx.lanl.gov/abs/1412.6000}{{\tt arXiv:1412.6000}}].

\end{thebibliography}\endgroup

\end{document}